\documentclass[twocolumn,english,american,prx, amsmath,amssymb, aps,superscriptaddress]{revtex4-2}
\usepackage[LGR,T1]{fontenc}
\usepackage{textcomp}
\usepackage[utf8]{inputenc}
\usepackage{amsmath}
\usepackage{graphicx}

\makeatletter

\DeclareRobustCommand{\greektext}{%
  \fontencoding{LGR}\selectfont\def\encodingdefault{LGR}}
\DeclareRobustCommand{\textgreek}[1]{\leavevmode{\greektext #1}}

\providecommand{\tabularnewline}{\\}

\makeatother

\usepackage{babel}
\begin{document}
\title{Active matter within flexible boundaries: a novel experimental approach
with \emph{T. aceti} nematodes}
\author{Christina M. Ceballos}
\affiliation{Department of Physics, California State University, Fullerton, Fullerton,
CA 92831, USA}
\author{\selectlanguage{english}%
Matthew S. Mizuhara}
\affiliation{\selectlanguage{english}%
Department of Mathematics and Statistics, The College of New Jersey,
Ewing, NJ 08628, USA}
\author{\selectlanguage{english}%
Mykhailo Potomkin}
\affiliation{\selectlanguage{english}%
Department of Mathematics, University of California, Riverside, Riverside,
CA 92521, USA}
\author{Anton Peshkov}
\affiliation{Department of Physics, California State University, Fullerton, Fullerton,
CA 92831, USA}
\begin{abstract}
We experimentally explore the collective behavior of the nematode
\emph{T. Aceti} inside flexible boundaries showing the emergence of
previously unreported states. These nematodes have been previously
shown to be able to synchronize their body oscillations in a favorable
condition of confined space. In this collective state, they are able
to exert a strong pushing force which we exploit to study how their
collective motion could deform a pliable boundary. Using a novel experimental
technique, we were able to simulate a non-rigid border at a liquid-liquid-gas
triple point. We discovered a state with periodic boundary deformations
which are due to the synchronized oscillations of the nematodes. More
interestingly, we found a novel state where the nematodes are able
to form multiple protrusions in the border which oscillate, rotate,
merge, and split. We propose a numerical model that is able to reproduce
some of the observed parameters of this state as well as offer an
analytical insight into protrusions dynamics. Our research opens a
pathway for exploring the interaction between active matter and flexible
interfaces with a new kinds of active matter agents forming dynamic
non-chaotic border deformations.
\end{abstract}
\maketitle

\section*{Introduction}

A lot of theoretical and experimental research have been dedicated
to the interaction of active matter with rigid borders, notably defining
the notion of active pressure \citep{takatori2014swim,solon2015pressure}
which in non-equilibrium active matter is a ``border dependent''
parameter \citep{nikola2016active}. However, the topic of interaction
of active matter with supple boundaries remains relatively unexplored,
especially experimentally. Investigations of active matter inside
rigid enclosures have shown the emergence of circulatory motion in
confined boundaries \citep{Deseigne2010,woodhouse2012spontaneous,wioland2013confinement,lushi2014fluid,bricard2015emergent}.
It is therefore possible to organize active matter by proper confinement
\citep{araujo2023steering}, and we could expect that the interaction
of organized particles with a pliable boundary could lead to interesting
new states. Indeed, numerical simulations have predicted a plethora
of possible states \citep{paoluzzi2016shape,wang2019shape,AbaurreaVelasco2019,Quillen2020,peterson2021vesicle,dhar2025_nema_drop_stability},
a lot of which still need to be observed experimentally. 

First experimental investigations of the interaction of active matter
with flexible walls were conducted with synthetic polar particles
such as vibrated active disks \citep{junot2017membrane} or Quincke
rollers \citep{kokot2022spontaneous} and has shown random non-equilibrium
deformations of the border. More recently experimental investigations
were conducted in 2-dimensional enclosures with both bacterial colonies
\citep{Xu2023} and microtubules \citep{Zhao2024,Sessa2026}. In both
cases the interaction between active-nematic particles and the soft
border produced short-lived periodic deformations of the border that
are due to internal topological defects of the active particles. Other
research concentrated on 3-dimensional droplets of microtubules \citep{sciortino2025active}
and bacteria \citep{Chang2026} where the chaotic bacterial flow induced
random deformations of the border. 

Contrary to the highly studied bacterial suspensions and microtubules,
that mostly produce chaotic ``turbulent'' motion, our nematodes
self-organize in confined spaces creating propagating metachronal
waves \citep{Peshkov2022}. In this collective state, the nematodes
produce both a very strong forward pushing force of up to $1\,\mu N$
\citep{Peshkov2022} and a strong fluid flow behind them \citep{Quillen2022}.
At a maximum packing concentration of around 30 nematodes per mm of
border, we can expect a pushing force of $30\,\mu N/mm$ comparable
in order of magnitude to the interfacial tension between water and
air $73\,\mu N/mm$\citep{Vargaftik1983}. It is therefore reasonable
to expect that our nematodes will be strong enough to deform the surface
of a sessile droplet that could be considered a flexible border. However,
when considering a droplet on a solid surface one must also take into
account the pinning of the edge to the surface \citep{Tadmor2021},
which may require much stronger initial forces to overcome. Therefore,
a droplet on a solid surface is not a good representation of a freely
moving boundary. To overcome this limitation, we developed a new technique
that allows us to study the border at liquid-liquid-gas triple point,
where pinning forces are absent and the droplet is completely free
to deform.

The three main discoveries of our research are. First, that we can
study the interaction between active matter and a flexible boundary
using the liquid-liquid-gas triple point as opposed to the previously
studied liquid-solid-gas\citep{kokot2022spontaneous,Xu2023} and liquid-liquid-solid\citep{Zhao2024,Sessa2026}
interfaces, where pinning is not negligible. Second, that we discovered
a state with periodic rippling of the border due to the metachronal
wave propagation. And finally, that we discovered a stunning state
where multiple dynamic protrusions form on the surface of the droplet,
that was not previously predicted in simulations or theory. We provide
an origin of this state by performing simulations and an analysis
of a simple model of active rods inside a flexible border.

\section*{Experimental methods}

\begin{figure}[h]
\includegraphics[width=1\columnwidth]{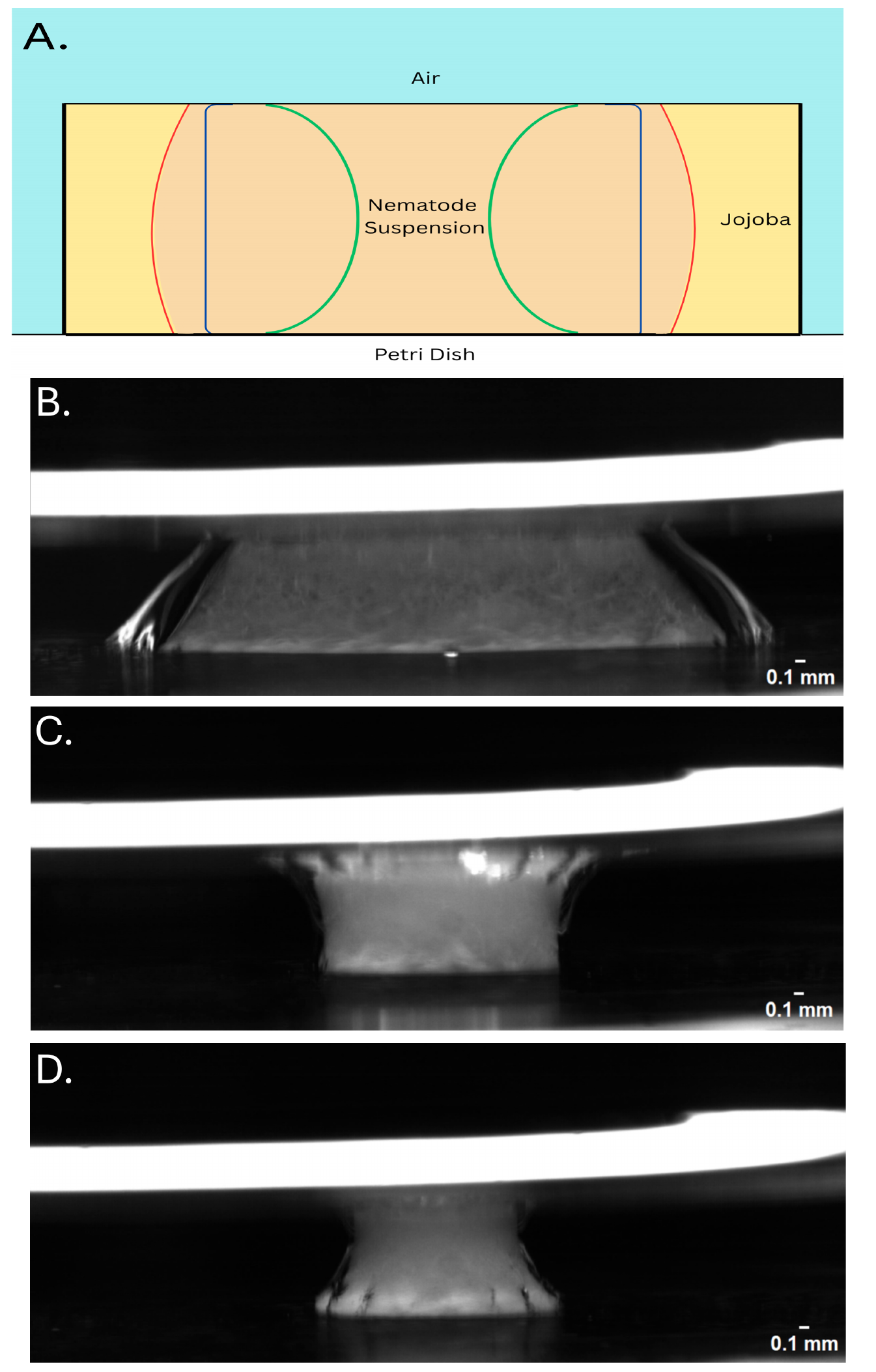}

\caption{a) Schematic of the experiment with the nematode droplet being extended
between the surface at the bottom and the air interface at the top.
Typical droplet shapes at the beginning, middle and end of experiment
are shown with respectively red, blue, and green outlines. Photos
b), c), and d) correspond to the same experiment at different time
intervals. b) Side photo of a droplet at the beginning of the experiment
when the Petri dish has just been filled with jojoba oil. c) Side
photo when periodic boundary deformation has formed at the top. d)
Side photo when long protrusions have formed.}\label{fig: schematic of the droplet}
\end{figure}

We prepare the nematodes for experiments by centrifuging the grow
medium with nematodes at 3000 rpm for 2 min. Depending on the desired
final density, the centrifugation process can be repeated several
times. We work with initial densities ranging from 20 to 250 nematodes
per ml ($n/ml$). Note that as the experiment progress, the droplet
will evaporate and the density therefore increase. We estimate the
``evaporated density'' of nematodes throughout the experiment by
taking into account the initial density and the calculated volume
of the droplet as detailed in the appendix.

We need to use another liquid to create the interface with the droplet
of nematodes. For this we employ \emph{Jojoba oil}, which is actually
not an oil (triglyceride) but a wax ester with a very high interfacial
tension with water reported in the range from $49\,\mu N/mm$ \citep{ijaz2021evaluation}\footnote{The specific gravity of jojoba oil reported in this article at 1.13
is very different from the standard industry measurements in the 0.85-0.9
range \citep{wisniak1987chemistry,spencer1988specifications,assaf2021jojoba},
casting doubts on the reported value of interfacial tension.} to as high as $70\,\mu N/mm$ \citep{Perillo2005}, the later value
measured for a monolayer film of jojoba on water. This interfacial
tension is much higher than the one of typical plant or mineral oils
and water $\sim20-30\,\mu N/mm$ \citep{fisher1985,gaonkar1989interfacial},
and is almost twice as high as the typical force produced by the nematodes
assuring that they will be able to deform the interface without breaking
it. 

A $200\,\mu l$ high density nematode droplet is deposited at the
bottom of a Petri dish whose surface was pretreated with a PDMS coating
(Rain-X). We then wait for the droplet to partially evaporate. Once
the droplet contact angle becomes small enough that a collective motion
with a metachronal wave is formed, we slowly start to fill the volume
of the Petri dish with jojoba oil, being careful not to fully cover
the nematode droplet with oil. During this process, the top of the
droplet with nematodes stays open to the air. Because of the affinity
of water to air, the droplet is being stretched between the solid
surface at the bottom and the air surface at the top as depicted on
Figure \ref{fig: schematic of the droplet} a). An easier to understand
analogy would be a droplet in air extended between two solid surfaces.
The initial contact angle both at the bottom and top is typically
above 90° as shown on Figure \ref{fig: schematic of the droplet}
b). The typical initial diameter of the droplet at the top is $10\,mm$.

However, because the top of the droplet remains open to the air, the
nematode droplet slowly evaporates over time, and the contact angle
both at the top air surface and the bottom Petri dish one decrease
as shown on Figure \ref{fig: schematic of the droplet} c) and d).
As demonstrated in our previous research \citep{Peshkov2022}, the
decrease in the contact angle leads to stronger synchronization between
the nematodes. As the synchronization between the nematodes increases,
the number of nematodes that our able to fit at the border increases
as well, which leads to an increasing force on the border as the experiment
progress. We record our experiments for up to 24 hours, by which time
the nematode droplet would typically fully disintegrate. Note that
the contact angle decreases both at the bottom solid surface and the
top air surface, and therefore the synchronization of nematodes happens
both at the top and the bottom. However, at the bottom solid surface
the droplet border is still subject to surface pining and is not free
to deform. Therefore, for the purpose of this study we neglect what
is happening at the bottom surface. We are performing our experimental
observation from the top and can clearly observe the deformation of
the top droplet border which is free to move.

\section*{Petal state}

\begin{figure*}
\includegraphics[width=1\textwidth]{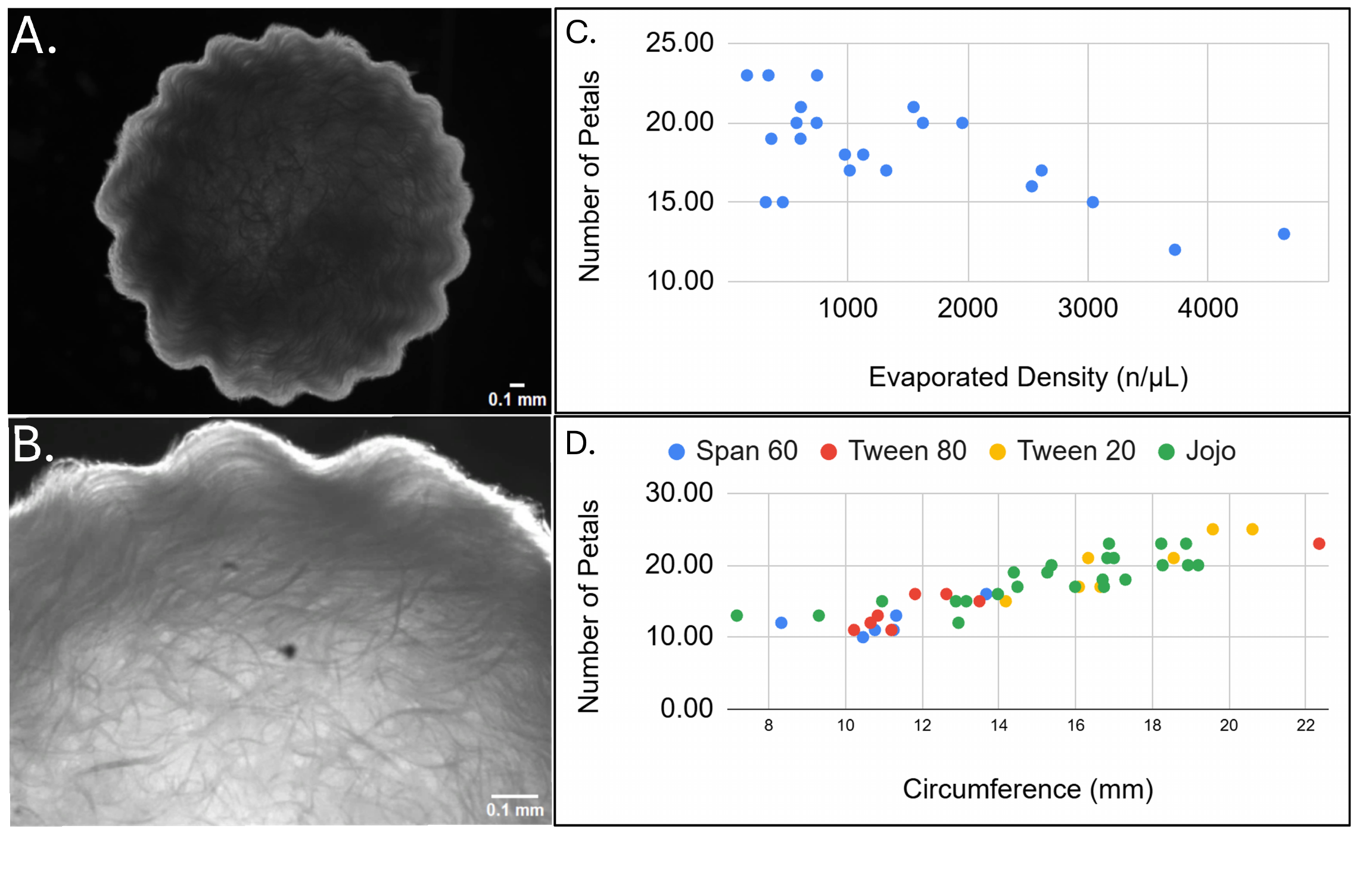}

\caption{a) Top photo of petal state. b) Zoom in on a few petals. c) Number
of petals as a function of nematode density. d) Number of petals as
a function of circumference.}\label{fig: Periodic deformations}
\end{figure*}

As the droplet evaporates and the synchronization of the nematodes
becomes greater, we observe the first reported collective state. In
this state a wave like deformation of the border occurs as depicted
in Figure \ref{fig: Periodic deformations} a) and supplemental video
1 \citep{Peshkov2026}. In this state the droplet acquires a ``flower''
like shape, we will therefore call it the ``petal state'' to avoid
confusion with the ``protrusion state'' described in the next section.
The petals are not stationary but rotate around the droplet. The mean
measured wavelength of the deformation is $0.86\,mm$. The mean amplitude
is not fixed and increases with time until reaching a maximum value
around $0.25\,mm$, as measured between the outermost part of the
petals and an imaginary circle formed by the innermost parts of the
petals. The typical wavelength of the petals as well as the amplitude
is similar to that observed in metachronal waves \citep{Quillen2021,Peshkov2022}.
Indeed, a zoom-in on one petal as shown in Figure \ref{fig: Periodic deformations}
b), illustrate that the petals correspond to the metachronal waves. 

When the droplet is circular, the nematode's bodies are on average
tilted at a low angle (\textasciitilde 30°) to the border, similar
to the case when the nematodes are in a fixed border sessile droplet
\citep{Quillen2021}. This means that when a nematode oscillates its
head, it will slightly push in and out on the border. As can be seen
in the supplemental video 2, the petals are due to this motion of
the nematodes head propagating with the metachronal wave. Note that
this leads to a slight asymmetry in the shape of the petals.

Figure \ref{fig: Periodic deformations} c) implies a slight negative
proportionality between the density of nematodes and the number of
deformations, implying that the pushing force does not control the
wavelength. Figure \ref{fig: Periodic deformations} d) show a linear
relationship between the circumference of a droplet and the number
of protrusions. This linear relationship is observed both for experiments
with jojoba oil and experiments where three different surfactants
were added to the oil to change the interface surface tension (surfactants
are described in the appendix). We see that adding the surfactants
has no effect on the wavelength of the perturbations, but just on
the diameter of the droplet, therefore the interfacial tension does
not control the parameters of this deformation. This further confirms
that these deformations are imposed by the oscillations of nematodes.

This state may look similar to the one predicted in numerical simulations
\citep{Quillen2020} of simple active particles interacting with a
flexible border where a wavelike deformation of the interface was
observed for some simulation parameters. The reason for such deformation
resides in shear forces applied by the active particles traveling
along the border creating pressure inhomogeneities \citep{nikola2016active}.
However, our deformations are driven purely by the synchronous oscillations
of our nematodes and the pressure gradient on the border is actually
different from the one predicted in \citep{nikola2016active} as explained
in the next section.

As for the experiments, the petal state may ressemble the periodic
deformations of the border previously reported for both bacterial
colonies\citep{Xu2023} and microtubules\citep{Zhao2024}, they are
actually fundamentally different. The deformation of borders for actin
filaments and bacteria colonies are transitory states driven by internal
topological defects. In these defects the bacteria or actin fibers
orient themselves perpendicularly to the border which drives the deformation.
Our state is in contrast long lasting and is due to metachronal wave
oscillations of the nematodes who are oriented closer to parallel
to the droplet circle.

\section*{Protrusions state}

\begin{figure*}
\includegraphics[width=1\textwidth]{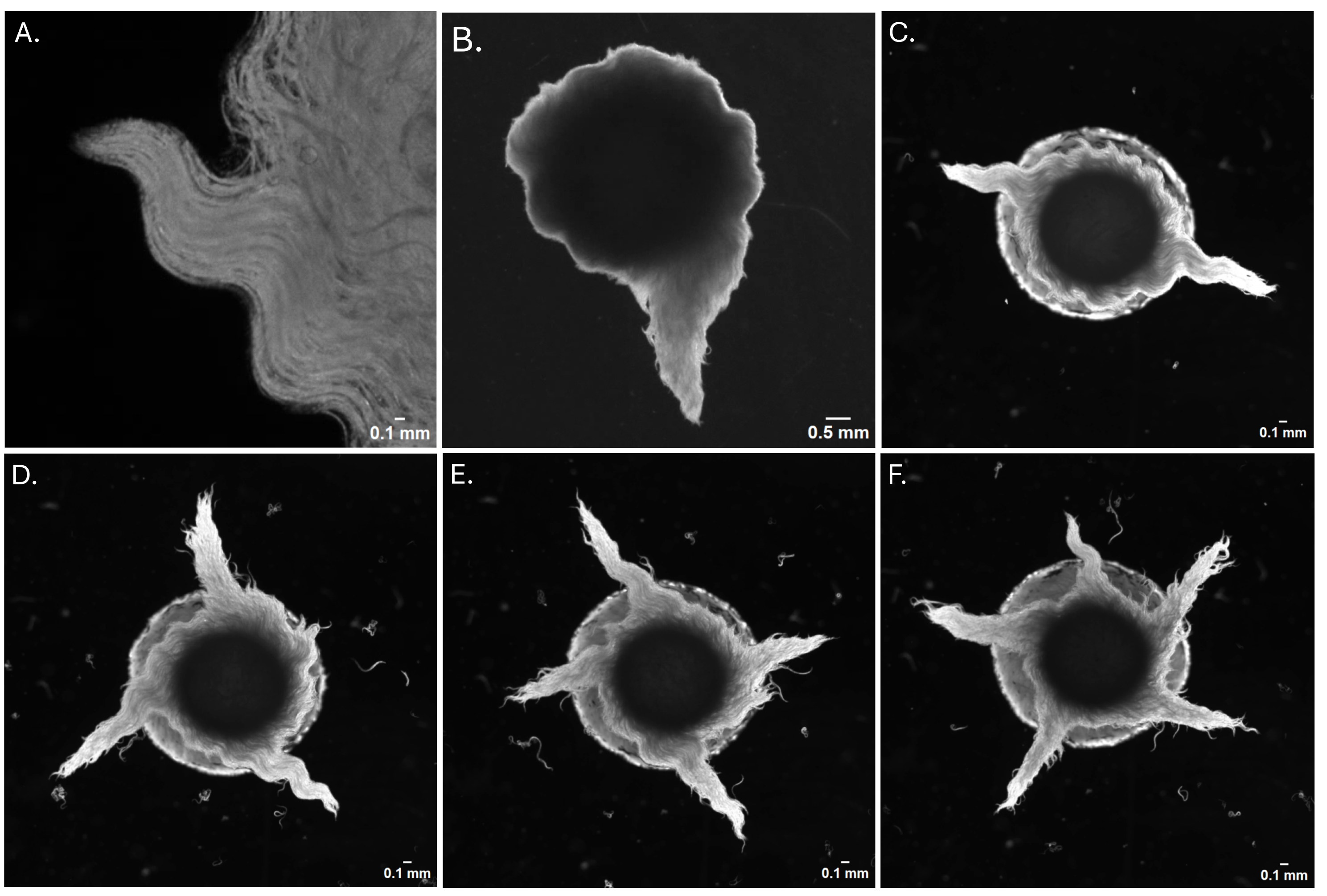}\caption{a) Initial formation of a protrusion from the side of a petal. b-f)
Photos of different numbers of protrusions. Note that by this time
in experiment the bottom of the nematode droplet in contact with the
surface is typically bigger in diameter than the top of the droplet
and can be seen on c-f) appearing as the bright circle underneath
the protrusions.}\label{fig:Photos of filaments}
\end{figure*}
As the droplet evaporates further, the synchronization of nematodes
becomes even greater as well as the forces exerted by them on the
border, which allow us to observe another collective state with long
protrusions. The theory of active pressure on curved walls predicts
that the greatest force should be applied at the ``tips'' of the
petals \citep{nikola2016active}. Therefore, one would expect that
as the pressure exerted by the nematodes increases, the petals would
deform at this point of greatest pressure. However, this is not what
we observe experimentally. As noted before, our nematodes are oriented
at a constant small angle toward a virtual circular surface. As the
real border has a petal shape, it means that at some point of the
petal, the angle between the nematodes and the border will be close
to zero, while at other points it will be close to 90°. We will therefore
expect that the points of greatest force on the wall will be located
where the nematodes are almost perpendicular to the border. 

\begin{figure*}[t]
\includegraphics[width=1\textwidth]{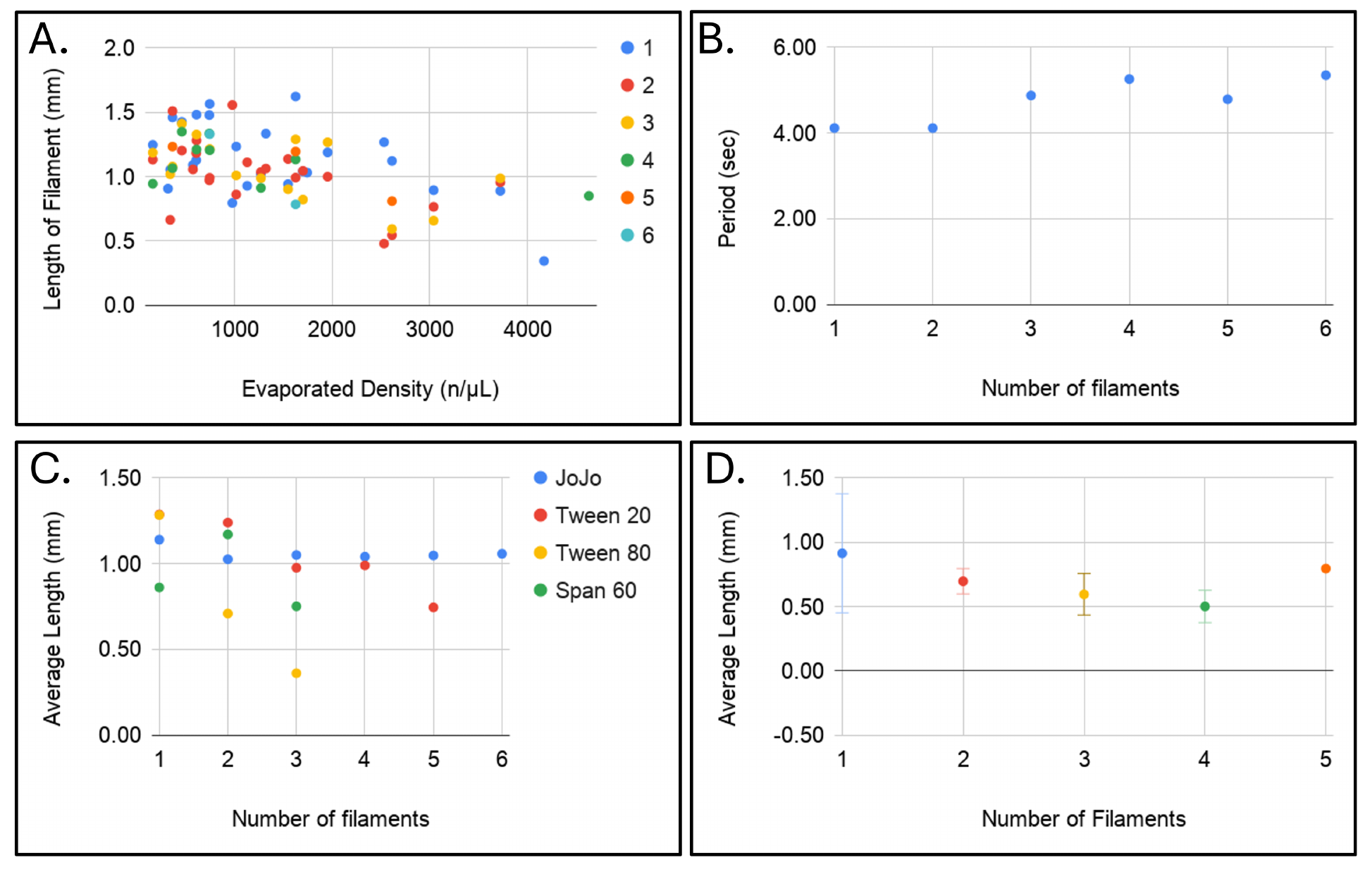}

\caption{a) Averaged time of a full rotation of a filament around the droplet
for different numbers of filaments. b) Length of protrusions as a
function of density for different number of protrusions. c) Average
length of filament vs number of filaments. d) Average period of oscillation
of protrusions in and out of the droplet as a function of the number
of filaments.}\label{fig:Protrusion parameters}
\end{figure*}

Indeed, when the synchronization of nematodes becomes strong enough,
we observe a formation of a long protrusion in one place of the border
at the location where the nematodes are perpendicular to the border.
Because of the tilt of the nematodes as compared to a circular surface,
the perpendicular location will not be at the end of the petals, but
rather along one of its sides. Therefore, the initial protrusions
are almost tangential to the droplet circle as seen on Figure \ref{fig:Photos of filaments}
a) and supplemental video 3. However, as the protrusions will grow,
their angle will become closer to perpendicular as compared to the
circular surface of the droplet. Interestingly, as the protrusion
will become more prominent in size and number, the ``petals'' will
become smaller until almost disappearing.

The number of protrusions typically increases over time; with a maximum
of six protrusions observed. When multiple protrusions are present,
they always space themselves symmetrically around the droplet to maximize
the distance between them as seen on Figure \ref{fig:Photos of filaments}
and supplemental video 4. The protrusions are not stationary, but
rotate in the opposite direction to the metachronal wave, which is
the direction of slow motion of the nematodes along the border. The
average time of a complete rotation around the droplet is dependent
both on the droplet size and the number of protrusions as shown on
Figure \ref{fig:Protrusion parameters} a).

The typical width of the protrusion is 0.75 mm, which would correspond
to about 25 nematodes diameters. The average length is about 1 mm
with a minimum of 0.5 mm and a maximum of 1.5 mm (Figure \ref{fig:Protrusion parameters}
b). This length has no clear dependence on the number of protrusions
and a weak dependence to the modification of surface tension with
surfactants as seen on Figure \ref{fig:Protrusion parameters} c).
The length of the protrusions is not stable in time as they oscillate
in and out of the droplet as seen in the supplemental video 4. Our
observations indicate that the protrusions shorten when the nematodes
at the front of the filament becomes temporally desynchronised, and
therefore the pushing force decreases. They regrow in length again
when the synchronization is restored. If the protrusion is not long
enough, the temporal loss of pushing force may make it completely
disappear. The typical period of oscillation in and out of the droplet
is around 4 to 6 seconds as shown on Figure \ref{fig:Protrusion parameters}
d).

The reason for these desynchronization events lies in overcrowding.
When a protrusion forms, it is almost impossible for any nematode
to escape from it. However, as the nematodes slowly drift along the
border, more and more of them will ``fall inside'' the protrusion.
At some time, the number of nematodes inside the protrusion will become
too big to maintain oscillations which would lead to a desynchronization.
This synchronization and desynchronization of nematodes are responsible
for the slow drift of protrusions along the border of the droplet
in the direction of the motion of nematodes which is also the direction
in which the protrusions are tilted. As can be seen in supplementary
video 3, when the nematodes desynchronize, a lot of them get injected
back into the droplet. The synchronized nematodes at the side of protrusion
in which it is tilted are able to bend the small interface and move
inside the protrusion. This leads to an effective displacement of
the protrusion in the direction in which it is tilted.

While these protrusions may look similar to the one recently reported
both in 2-dimensional \citep{Xu2023} and 3-dimensional bacterial
droplets \citep{Chang2026}, they are in fact different states. Our
protrusions are long lived in time and are equally spaced around the
droplet. The protrusions observed in bacterial colonies are ``escaping''
out of the droplet, are randomly located, and can form intersecting
networks. We observe the formation of the same type of random escaping
protrusions by the end of the experiment when the contact angle becomes
very small, and the droplet is being ``thorn apart'' by escaping
nematodes as seen in the appendix Figure \ref{fig:Meadowfoam} and
supplemental video 5. Note that the disintegration of the droplet
at high internal pressure was numerically predicted earlier \citep{diaz2024active}.
Interestingly, we find that in all our experiments the droplet will
disintegrate when reaching exactly the same reduced diameter of about
$5\,mm$. Indeed, as the droplet diameter decreases, we expect the
internal Laplace pressure to increase. This circumstance, combined
with the active force exerted by the nematodes, could cause the border
to break. We further study the ``escaping'' style protrusions in
a separate upcoming work \citep{Nguyen} where we show that the surface
breakage always happens at a given surface curvature.

\section*{Simulations and theory}

\subsection*{Model}

To elucidate the formation of the protrusions, we developed a numerical
simulation model of active rods inside a flexible boundary based on
the model studied in \citep{brown2025boundary}. The main assumptions
of the model are as follows: 
\selectlanguage{english}%
\begin{enumerate}
\item Nematodes are represented as self-propelled rods obeying overdamped
dynamics. 
\item Nematodes interact sterically, preventing overlaps. 
\item The droplet boundary is modeled as an elastic curve undergoing active
forces exerted by nematodes accumulated at the boundary. 
\item In turn, contact with the droplet boundary generates a torque on each
nematode, reorienting it so that it forms a tilting angle depending
on the local curvature of the boundary. 
\end{enumerate}
The state of the $i$th nematode is thus described by a location $\boldsymbol{r}_{i}=(x_{i},y_{i})$
and orientation $\varphi_{i}$ obeying \foreignlanguage{american}{
\begin{equation}
\left\{ \begin{array}{l}
\xi_{s}\dot{\boldsymbol{r}}_{i}=\boldsymbol{F}_{\text{prop},i}+\boldsymbol{F}_{\text{st},i}+\boldsymbol{F}_{\text{dr},i}\\
\xi_{o}\dot{\varphi}_{i}=\Omega_{\text{st},i}+\Omega_{\text{dr,i}},
\end{array}\right.\label{eq:nematode}
\end{equation}
}with force contributions generated from intrinsic propulsion (prop),
steric interactions (st) and droplet interactions (dr). On the other
hand, the droplet is represented by a discretization consisting of
$N_{d}$ nodes: $\boldsymbol{R}_{i}:=(x^{(d)}_{i}(t),y^{(d)}_{i}(t)),\;i=1,\dots,N_{d}$
evolving by \foreignlanguage{american}{
\begin{equation}
m\ddot{\boldsymbol{R}}_{i}=\boldsymbol{F}_{\text{bend},i}+\boldsymbol{F}_{\text{spring},i}+\boldsymbol{F}_{\text{damp},i}+\boldsymbol{F}_{\text{active},i}.\label{eq:droplet}
\end{equation}
}The boundary experiences passive elastic forces from bending (bend),
spring (spring), and damping (damp), along with active contributions
from nematode interactions (active). Details of the model and parameter
values for all simulations results that follows are available in the
Appendix. While other meaningful physical effects such as nematode
synchronization have been omitted for simplicity, our simulations
reveal that even the minimal interactions above are sufficient to
generate a wide range of dynamical behaviors.

\begin{figure*}
\includegraphics[width=0.6\textwidth]{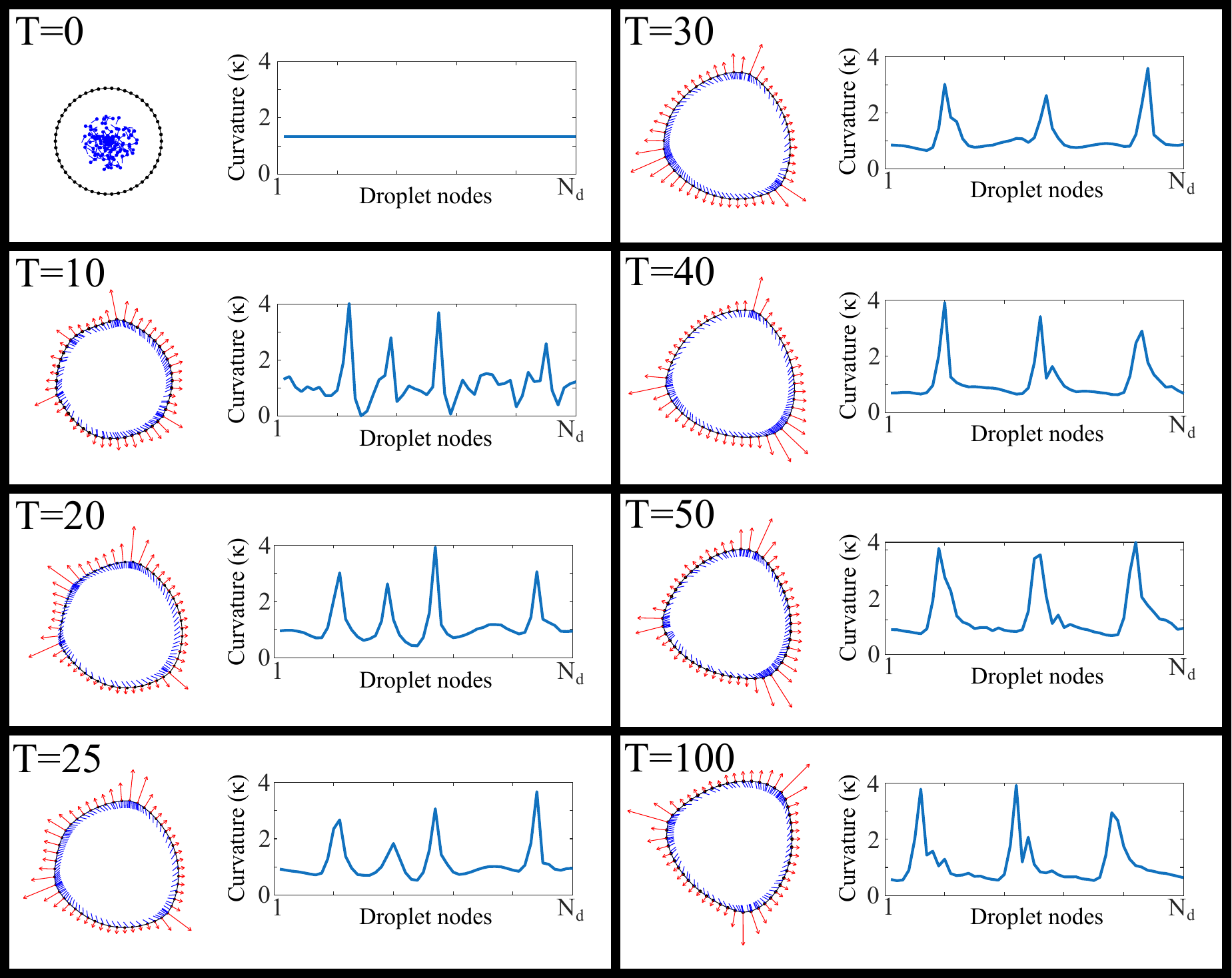} \caption{\textbf{Numerical example exhibiting protrusion formation and merging.}
Sub-figures represent snapshots at various points in time together
with the curvature profile. Nematodes are represented as blue rods
with a thicker node at the head, and droplet nodes are represented
as black segments. Red segments show the magnitude of local forces
generated by nematodes on the boundary. Initially ($T=0$), all nematodes
are randomly generated within the droplet and they quickly accumulate
to the boundary. From $T=20$ to $T=30$, one can initially observe
four protrusions which merge until three equidistant protrusions remain.
These protrusions rotate clockwise with a steady angular speed (opposite
the direction nematodes swim along the droplet boundary).}
\label{fig:numerical_results_1}
\end{figure*}

\subsection*{Results}

\emph{Feedback between nematode clusters and boundary curvature generates
protrusions.} Nematodes are initialized randomly within a circular
droplet. They quickly converge to the droplet boundary and they cover
it entirely. Due to nematode tilting angle with the boundary, they
also swim along the boundary. For ease of clarity, in our implementation,
we always enforce nematodes to swim counterclockwise along the droplet
boundary.

\begin{figure*}[t]
\includegraphics[width=0.8\textwidth]{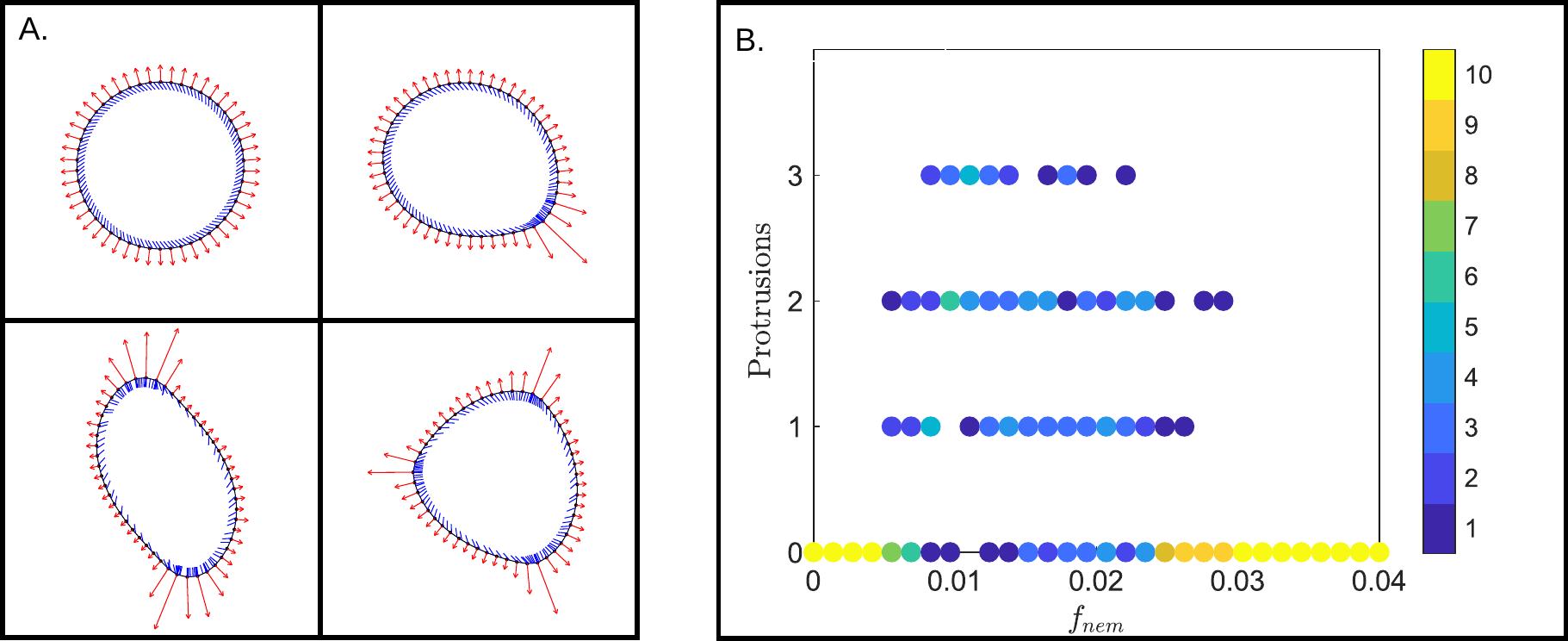} \caption{\textbf{Illustration of multi-stability suggested by numerical observations.}
We performed 10 numerical simulations for each value of the nematode
pushing force $f_{\mathrm{nem}}$ using random initial conditions
(rod locations and orientations). The snapshots on the left show representative
snapshots of four stable behaviors (for the same fixed-parameter values).
The graph on the right shows how many times we observed zero, one,
two, or three stable protrusions at large times among the 10 simulations
for each value of $f_{\mathrm{nem}}$. The color of each data point
indicates the number of occurrences of the corresponding protrusion
count for a given value of $f_{\mathrm{nem}}$, as shown by the colorbar
on the right.}
\label{fig: numerical_results_2}
\end{figure*}

Small perturbations of the nematode density around the boundary lead
to symmetry breaking and the formation of protrusions. This instability
appears to result from a positive feedback mechanism between nematode
clustering and deformations of the boundary. Specifically, as nematodes
cluster, their collective forces locally deform the boundary, increasing
its local curvature. The increased curvature, in turn, slows the motion
of nematodes leading to further growth of clusters. The results of
our simulations are presented in Figures \ref{fig:numerical_results_1}
and \ref{fig: numerical_results_2}.

\emph{Protrusions are equally spaced and drift.} We note that nematode
clusters are dynamic: nematodes constantly merge from one side of
the protrusion and escape from the other. The protrusions themselves
drift in the opposite direction that nematodes swim (i.e., the protrusions
drift clockwise). This can be understood as a similar mechanism to
shock wave propagation in traffic flows: as new nematodes enter the
cluster new protrusive forces form in the clockwise direction. Likewise,
as old nematodes exit the cluster, the protrusive forces decrease
on the counterclockwise side of the protrusion. Together this results
in a drift in the opposite direction of motion of nematode swimming
which is contrary to what we observe in experiments where the drifting
of protrusions is in the direction of nematodes motion. We suggest
that the reason for this discrepancy is that the experimental protrusion
displacement is due to the nematode synchronization and desynchronization
events which we do not simulate in the current model. Figure \ref{fig:numerical_results_1}
provides representative snapshots demonstrating protrusion formation
and drifting. Similar to experimental findings, after transience,
we observed that if multiple protrusions were present, then they were
equally spaced around the droplet. 

\emph{Protrusions exhibit merging events and multistability.} Once
protrusions form, there can be interesting transient behaviors wherein
protrusions move relative to each other and merge. For example, Figure
\ref{fig:numerical_results_1} shows a representative such simulation:
at around $T=20$ four protrusions form. By $T=30$, two of these
protrusions merge to result in three protrusions. These three protrusions
persist until the simulation ends, suggesting that this is the stable
configuration for these initial data. Protrusion merging is also observed
in experiments.

Surprisingly, even upon fixing physical parameters, the number of
protrusions in the final stable configuration of the system varied.
In our simulations we observed that zero, one, two, and three protrusion
configurations could all be stable, depending solely on the random
initialization (see supplemental videos (6-9)). This multi-stability
was robust over a range of nematode protrusion forces, see Figure
\ref{fig: numerical_results_2}. It is interesting to note that protrusions
will form only for a range of forces, with both too small and too
big forces leading to a circular droplet, which is in agreement with
previous simulations of active particles inside soft borders \citep{Quillen2020}. 

\selectlanguage{american}%

\section*{Discussion and Conclusions}

Both our simulations and previous research \citep{nikola2016active,Quillen2020}
has shown that producing stable deformations such as periodic petals
and long protrusions, requires to reach a \emph{sweet spot} between
the force applied by the active agents and the rigidity of the border.
If the force is too low, the active particles would not be able to
deform the border. If the force is too high, the active particles
will apply a very high uniform \emph{active pressure}, that will make
the droplet surface smooth again. We tried to repeat this experiment
with several different plant and petroleum derived oils, whose interfacial
tension with water is low, as well as liquids such as squalene and
squalane whose interfacial tension with water is larger in the $45-47\,\mu N/mm$
range \citep{hildebrandt2016squalene}. Only jojoba oil gave us consistent
repeatable results (see appendix) in producing the protrusions state,
showing that it's the only liquid which gives the right ratio between
the interfacial tension and the forces produced by the nematodes.
Note that when talking about the rigidity of the border, one could
argue about the necessity of taking into account the bending energy
of the interface defined by the bending rigidity $\kappa$ and not
just the surface tension $\gamma$. However, the typical elastocapillary
length $l_{ec}=\sqrt{\kappa/\gamma}$, defining the length scale below
which bending rigidity becomes important, is of the order of nanometers,
many orders of magnitude smaller than the size of our protrusions.
Therefore, it can safely be neglected for our purposes. 

The main difficulty of working with jojoba oil is that being a natural
plant derived product, its physical properties vary from batch to
batch, which makes obtaining conclusive quantitative values difficult.
Further experiments will need to explore the possibility of substituting
jojoba oil with a synthetic liquid with well-controlled physical properties.
The only well-known chemical analog to jojoba oil, a liquid wax ester,
would be the \emph{sperm oil}; liquid obtained from sperm whales whose
harvest and sell are internationally prohibited since 1987. Other
marine animals derived wax esters such as \emph{orange roughy oil}
and \emph{copepod oil} exist but will be subject to the same batch
variability as jojoba oil and are hard to procure. While synthetic
\emph{esther lubricants} were developed as a substitute for sperm
oil for industrial applications, their interfacial tension with water
is actually low in the order of $\sim30\,\mu N/mm$ \citep{CoelhodeSousaMarques2021}.
This means that a totally different class of liquids will have to
be explored as a substitution. Possible candidates could include various
liquid \emph{n-alkanes} that have a water interfacial tension in the
range of $50-54\,\mu N/mm$ at room temperatures \citep{zeppieri2001alkane_tension}.

It is well know that cell locomotion and swimming is achieved through
dynamical shape change \citep{lauga2009hydrodynamics,wu2016amoeboid,noselli2019swimming}
and it would be interesting to achieve such motion with droplets enclosing
active particles. Past experimental research has shown the possibility
of creating a random displacement of droplets enclosing active matter
\citep{ramos2020bacteria,kokot2022spontaneous}. But no research so
far has been able to produce controlled motion in one direction, except
when such direction was imposed by external conditions \citep{rajabi2021directional}.
It has been proposed that a directional locomotion of active droplets
could be achieved from the interaction of a droplet shape deformation
and the surrounding fluid \citep{kawakami2025active,kawakami2025migration}.
In our current setup, the bottom of the droplet is pinned to the surface
preventing it from having large displacements. However, we were able
to observe states when the droplet was able to rotate around the bottom
``anchor'' (see supplemental video 10). It will be interesting to
explore the possibility of creating a free-moving droplet of our nematodes
on top of another liquid, which could open the path of controlled
droplet motion given the non-chaotic nature of the motion of our nematodes. 

While the emergence of protrusions can be explained by an instability
of the state in which accumulated swimmers are uniformly distributed
along the boundary, the tendency of the protrusions to become equally
spaced is more intriguing. Such a distribution of protrusions could
be obtained by imposing a long-range repulsion between them, causing
them to maximize the distances between one another, as observed in
both experiments and simulations. However, there is no apparent physical
reason for the protrusions to repel, and such a repulsive force would
prevent them from merging, contrary to what we observe in simulations.
Instead, to explain this phenomenon, we consider a simplified physical
model consisting of an elastic closed curve whose equilibrium shape
is circular. We assume that active forces of equal magnitude are applied
to the boundary at locations that are free to vary. Here, these active
forces represent protrusions or, equivalently, clusters of nematodes
accumulated at the droplet boundary. It turns out (see the Appendix
for details) that the elastic energy is minimized when the points
of the application of these forces are equally spaced. This mathematical
observation suggests that the tendency of protrusions to become equally
spaced may arise naturally from the elastic response of the droplet
boundary to the active pressure exerted by nematode clusters. 

Although the active rod model successfully captures the formation
of protrusions, their shape is much smoother than that observed in
experiments. The simulated protrusions have a smooth shape, whereas
the experimental protrusions are reminiscent of tongues of flame,
with a sharp kink at the tip. Capturing the experimentally observed
protrusions will require a more detailed model of individual nematodes
and the possibility for their synchronization. The absence of nematodes
synchronization in the numerical model is notably the reason why we
obtain a drift of protrusions in opposite directions in experiments
and simulations as explained previously. We anticipate that synchronized
swimming of neighboring nematodes increases the local active pressure
exerted on the droplet boundary and thus affects the dynamics of protrusion
formation and motion.

In conclusion, we were able to design a new experimental approach
that allows us to study the interaction of active matter with a soft
border at a liquid-liquid-gas triple point which allows for a completely
free deformation contrary to experiments that try to deform a border
pinned to a solid surface. Because our nematodes are able to synchronize
their oscillations to produce moving metachronal wave, we were able
to discover two new states: one in which the metachronal wave produces
a petal-like deformation of the border, and one where an increased
active pressure leads to the formation of long dynamic protrusions
at the border of the droplet. The easiness of growth and experimentation
with our nematodes together with the strong forces they produce make
them a very interesting active system to study. Indeed, the synchronized
nematodes not only produce a forward pushing force but also a backward
flow of liquid \citep{Quillen2022}. We exploit this fact in our sister
article \citep{Robinson2026}, where we demonstrate the possibility
of creating on demand fluid flows by adjusting the synchronization
of our nematodes.

\section*{Author contributions}

Christina Ceballos and Anton Peshkov conceived the study and designed
the experiments. Christina Ceballos performed the experiments and
analyzed the experimental data. Matthew Mizuhara and Mykhailo Potomkin
developed the numerical and analytical model, performed the simulations,
and analyzed the numerical results. All authors edited the manuscript
and gave their approval for the final publication.

\section*{Data and software availability}

All analyzed data supporting the presented graphs will be deposited
on Figshare \citep{Peshkov2026} and made publicly available upon
article acceptance. The code used for numerical simulations is currently
available on Zenodo \citep{Mizuhara2026}. Original video recordings
of experiments are not shared due to their size, but can be individually
requested from the authors. 

\section*{AI use}

AI tools (ChatGPT and Claude) were used to help with bibliographical
research for the manuscript. No AI tools were used to design any parts
of the experiment, for data analysis or interpretation, code writing,
figure preparation or manuscript text editing.

\section*{Funding}

The work of Christina Ceballos and Anton Peshkov has been supported
by NSF grant number PHY-2412690 and an internal California State University,
Fullerton Junior/Senior grant. Mykhailo Potomkin was supported by
the UC Riverside Regents Faculty Fellowships grant. Matthew S. Mizuhara
acknowledges support from NSF grant DMS-2406942 and use of the ELSA
high performance computing cluster at The College of New Jersey, which
was funded in part by NSF grants OAC-1826915 and OAC-2320244.

\section*{Conflict of interests}

The authors have no competing interests to declare.

\bibliographystyle{apsrev4-2}
\bibliography{nematodes_elastic_borders_references_new}

\section*{Appendix}

\subsection*{Experimental methods}

\subsubsection*{Grow}

We grow our nematodes in 1:1 solution of water and apple cider vinegar
at $5\%$ acidity inside $250\,ml$ cell culture flasks. We complement
it with slices of apple serving as a food source. The solution is
initially kept at room temperature for a few weeks to reach peak density
of nematodes. It is then moved to an incubator at a temperature of
$15{^\circ}C$ to maximize the nematodes lifespan. The reproduction
cycle of the nematodes takes several days, and their life span can
reach several months. Therefore, there is no significant variation
in the number of live nematodes during the 24-hours maximum duration
of our experiments.

\subsubsection*{Density computation}

\selectlanguage{english}%
To determine the initial density of nematodes, we diluted the dense
nematode suspension 10-50 times in a 50:1 solution of water and glycerol
to prevent droplet diameter shrinkage during evaporation. 10 \textgreek{μ}l
droplets of the diluted solution were placed on a glass slide and
evaporated. The number of nematodes in each droplet was then manually
counted under a microscope. This provides the mean and the error estimate
for the initial density.

\selectlanguage{american}%
Because the droplet evaporates over time, the density of nematodes
changes accordingly. To estimate the ``evaporated density'', we
use the side images of our droplet to obtain the ``profile'' of
the droplet in Image J. From this side profile we can then obtain
the volume of the whole droplet at a time $t$ by performing a 360°
revolution. By comparing this volume to the initial volume of the
droplet of $100\,\mu l$, we can then easily obtain the evaporated
density. 

\subsubsection*{Surfactants}

To investigate the effect of surface tension on the presented state,
we used several surfactants that we added to the jojoba oil. The used
surfactants were Tween 20, Tween 80, and Span 60. The percentage densities
that we tried ranged from 0.0025\% to 1\% by volume. Tween 20 and
80 are liquid and were mixed directly with a vortex machine. Span
60 is solid. To incorporate it into oil, we heated the latter to $80{^\circ}C$
and mixed it with a magnetic stirrer at $450\,rpm$ for 10 minutes,
until all the surfactant particles become dissolved. There is no way
to easily estimate the reduction in surface tension from the added
surfactant and the reduction is not forcefully proportional to the
percentage of added surfactant. Therefore, we only report the surfactant
results ``in bulk'', mainly to show that it has no or minimal effect
on the results. Note that adding the surfactant directly to the droplet
with nematodes led to the quick destruction of the droplet by nematodes
without the characteristic ``petal'' and ``protrusions'' states
described in the article.

\subsection*{Alternative oils}

As discussed in the main text of the article, most natural and synthetic
oils have a relatively low interfacial tension with water of $20-30\,\mu N/mm$.
We tried to repeat our experiment with mineral oils of different density,
but in all cases were not able to observe the protrusion state, as
the nematodes will destroy the droplet well before reaching that state.
The same with squalane, which has a reported interfacial tension of
$\sim45\,\mu N/mm$ with water. The only partial success that we had
was with meadowfoam oil. While being a vegetable oil (triglyceride)
obtained from the seeds of \emph{Limnanthes alba} and not a wax ester
as jojoba, its fatty acid profile is dominated by very long-chain
monounsaturated acids \citep{Zielinska2020meadowfoam} similar to
that of jojoba and has a small amount of wax esters \citep{Isbell1996}.
In cosmetic use, it is considered a possible substitute for jojoba
oil. Unfortunately, there is no known measure of the interfacial tension
between meadowfoam oil and water but based on our results we estimated
it to be much higher than the one typical of vegetable oils. 

In our experiments with meadowfoam oil, we were able to observe the
petal state as well as an initial stage of the protrusion state. In
experiments with jojoba, we observed protrusions that oscillate in
and out of the droplet as the nematodes synchronize and desynchronize.
In contrast, in meadowfoam oil we observed the initial formation of
a protrusion. However, as the protrusion will grow and the nematodes
in it become more numerous, the nematodes at the front will be able
to break the oil-water interface and escape into the oil, which will
effectively destroy the protrusion. This can be seen in Figure \ref{fig:Meadowfoam}
as well as in the supplemental video 5. 

\begin{figure}
\includegraphics[width=1\columnwidth]{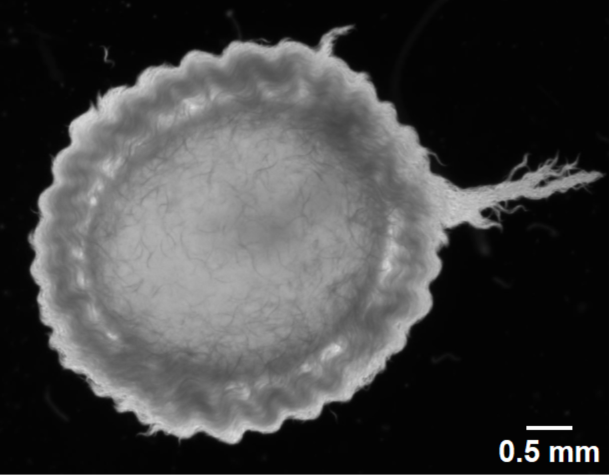}

\caption{Protrusion state in meadowfoam oil. Nematodes can be seen escaping
from a protrusion.}\label{fig:Meadowfoam}

\end{figure}

\subsection*{Description of computational model}

The state of $i^{\mathrm{th}}$ nematode is described by the center
location $\boldsymbol{r}_{i}=(x_{i},y_{i})$ and the orientation angle
$\varphi_{i}$ with respect to $x$-axis. The number of nematodes
will be denoted below by $N_{f}$. An individual nematode is propelled
towards its front with the force $\boldsymbol{F}_{\text{prop},i}=f_{\text{prop}}\boldsymbol{p}(\varphi_{i})$,
where $\boldsymbol{p}(\varphi_{i})=(\cos(\varphi_{i}),\sin(\varphi_{i}))$
is the unit orientation vector. To describe steric interactions between
nematodes, that is, that nematodes tend not to overlap, we compute
the overlap function between two nematodes represented as a series
of disks of radius $r$ with centers uniformly distributed along the
corresponding rod segment: $\boldsymbol{r}_{i}+(2k-1)/(2N_{n})\ell\boldsymbol{p}(\varphi_{i})$,
where $k=1,..,N_{n}$, $N_{n}$ is the number of sub-disks, and $\ell$
is the nematode's length. The overlap area between two sub-disks of
radius $r$, whose centers are separated by a distance $d$, is given
by 
\[
\mathcal{A}=r^{2}\left(2\psi-\sin(2\psi)\right),\quad\text{where}\quad\psi=\arccos\!\left(\frac{d}{2r}\right).
\]

\begin{figure}[t]
\includegraphics[width=0.46\textwidth]{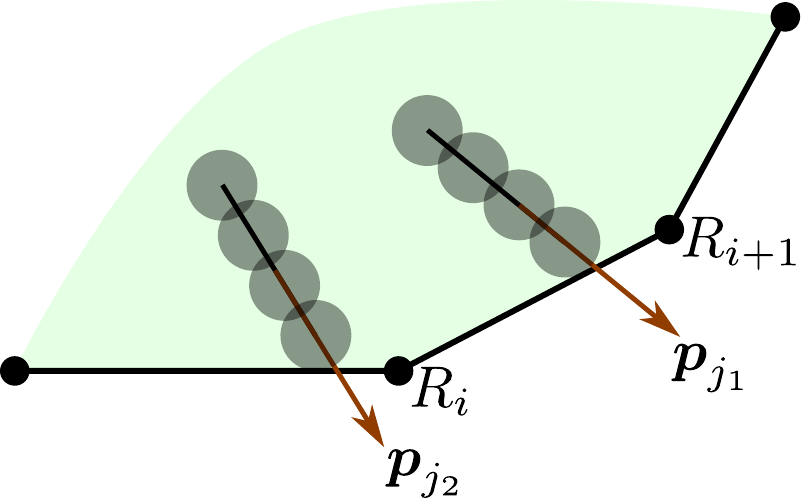} \caption{Model schematics. The illustration depicts two rods composed of 4
disks each. Both rods, labeled $j_{1}$ and $j_{2}$, attach to the
boundary of the droplet.}
\end{figure}

If we denote the overlap of the $k^{\mathrm{th}}$ and $\ell^{\mathrm{th}}$
disks of the $i^{\mathrm{th}}$ and $j^{\mathrm{th}}$ nematodes by
$\mathcal{A}_{kl}$, then the full overlap between two nematodes,
indexed by $i$ and $j$, is computed by 
\[
U_{ij}=\sum\limits^{N_{n}}_{k,l=1}\mathcal{A}_{kl}.
\]
We naturally impose the penalization of overlaps between sub-disks
of distinct nematodes, thus defining the steric force and torque as
follows: 
\[
\boldsymbol{F}_{\text{st},i}=-\dfrac{\partial}{\partial\boldsymbol{r}_{i}}\left[\sum\limits_{j\neq i}U_{ij}\right]\text{ and }\Omega_{\text{st},i}=\dfrac{\partial}{\partial\varphi_{i}}\left[\sum\limits_{j\neq i}U_{ij}\right].
\]
Dynamics of nematodes, swimming inside the droplet and not interacting
with its boundary, is given by system~(\ref{eq:nematode}). In this
system, $\xi_{s}$ and $\xi_{o}$ are damping coefficients. The last
term in the first equation, $\boldsymbol{F}_{\text{dr},i}$, enforces
that the nematode cannot escape the droplet. To this end, one can
define the reaction force $\boldsymbol{F}_{\text{dr},i}$ exerted
by the droplet boundary on the nematode, touching the boundary, as
follows 
\[
\boldsymbol{F}_{\text{dr},i}=-\boldsymbol{\nu}^{\text{T}}\boldsymbol{\nu}\left(\boldsymbol{F}_{\text{prop},i}+\boldsymbol{F}_{\text{st},i}\right).
\]
Here, $\boldsymbol{\nu}=(\cos(\varphi_{\nu}),\sin(\varphi_{\nu}))$
denotes the outward normal vector to the droplet boundary at the point
where the $i^{\text{th}}$ nematode makes contact with the boundary.
One can readily verify that this term causes the nematode to swim
tangentially, that is, along the droplet boundary. However, in the
numerical implementation, due to the discrete time step, a nematode
swimming away from the wall is more likely to be moved outside the
droplet rather than descend exactly onto the droplet boundary. If
the nematode is moved outside the droplet, we adjust its position
by relocating it onto the nearest point on the boundary. To define
the boundary reaction torque $\Omega_{\text{dr,i}}$, we note that
the torque generated by the force $\boldsymbol{F}_{\text{dr},i})$
is given by $(\boldsymbol{p}_{i}\times\boldsymbol{F}_{\text{dr},i})\cdot\boldsymbol{e}_{z}$.

\smallskip{}

The droplet boundary is described by $N_{d}$ nodes: $\boldsymbol{R}_{i}:=(x^{(d)}_{i}(t),y^{(d)}_{i}(t)),\;i=1,\dots,N_{d}$.
Initially, the droplet boundary is taken to be a regular polygon with
$N_{d}$ vertices inscribed in a circle of radius $R$. The distance
between neighboring vertices at $t=0$ is therefore $\Delta s=\frac{2\pi R}{N_{d}}.$
Then, the droplet evolves by equation (\ref{eq:droplet}). The bending
force $\boldsymbol{F}_{\text{bend},i}$ is given by considering nodes
as connected by thin elastic beams. Then 
\[
\boldsymbol{F}_{\text{bend},i}=\frac{-\alpha_{\text{bend}}}{(\Delta s)^{3}}({\boldsymbol{R}}_{i-2}-4{\boldsymbol{R}}_{i-1}+6{\boldsymbol{R}}_{i}-4{\boldsymbol{R}}_{i+1}+{\boldsymbol{R}}_{i+2}).
\]
Here and below, the index is taken modulo $N_{d}$, i.e., $\boldsymbol{R}_{i}=\boldsymbol{R}_{\hat{i}}\quad\text{whenever}\quad i\equiv\hat{i}\pmod{N_{d}}.$
The linear spring force is $\boldsymbol{F}_{\text{spring},i}=-\dfrac{\partial\mathcal{E}_{d}}{\partial\boldsymbol{R}_{i}}$,
where the total elastic energy of the droplet boundary is 
\[
\mathcal{E}_{d}=\dfrac{k_{s}}{2}\sum\limits^{N_{d}}_{i=1}(\|\boldsymbol{R}_{i}-\boldsymbol{R}_{i-1}\|-\Delta s)^{2}.
\]
The viscous damping force is given by $\boldsymbol{F}_{\text{damp},i}=-m\gamma_{damp}\dot{{\bf x}}_{i}.$
The active force $\boldsymbol{F}_{\mathrm{active},i}$ arises from
the pushing of the boundary by accumulated nematodes. If the front
of the $k$th nematode, denoted by $\boldsymbol{r}_{f,k}$, lies between
boundary nodes $\boldsymbol{R}_{i}$ and $\boldsymbol{R}_{j}$ (with
$j=i\pm1$), then the nematode exerts a force on each node proportional
to its distance to the other node. Specifically, the force on boundary
node $\boldsymbol{R}_{i}$ is given by 
\[
\boldsymbol{F}_{i,k}=f_{\mathrm{nem}}\frac{\|\boldsymbol{R}_{j}-\boldsymbol{r}_{f,k}\|}{\|\boldsymbol{R}_{j}-\boldsymbol{R}_{i}\|}\,\boldsymbol{N}_{i},
\]
where $\boldsymbol{N}_{i}$ is the unit normal at node $\boldsymbol{R}_{i}$,
orthogonal to the segment $[\boldsymbol{R}_{i},\boldsymbol{R}_{j}]$
and pointing outward from the droplet. Finally, the term 
\[
\boldsymbol{F}_{\mathrm{active},i}=\sum^{N_{n}}_{k=1}\boldsymbol{F}_{i,k}.
\]
is the total force exerted by accumulated nematodes on the node $i$.

\begin{table}
\begin{ruledtabular}
\begin{tabular*}{1\columnwidth}{@{\extracolsep{\fill}}|c|c|c|}
\hline 
\foreignlanguage{english}{Parameter } & \foreignlanguage{english}{Description } & \foreignlanguage{english}{Non-dim. Value }\tabularnewline
\hline 
\hline 
$N_{nem}$ & Number of nematodes & 125\tabularnewline
\hline 
\foreignlanguage{english}{$N_{\text{d}}$ } & Number of droplet nodes & 50\tabularnewline
\hline 
\foreignlanguage{english}{$R_{\text{drop}}$ } & Radius of initial droplet & 0.75\tabularnewline
\hline 
\foreignlanguage{english}{$k_{\text{spring}}$ } & \foreignlanguage{english}{Hooke's spring constant} & 7.0\tabularnewline
\hline 
\foreignlanguage{english}{$k_{\text{ang}}$ } & Bending constant & $10^{-6}$\tabularnewline
\hline 
\foreignlanguage{english}{$\gamma_{\text{damp}}$ } & Damping coefficient of nodes & 1\tabularnewline
\hline 
\foreignlanguage{english}{$\ell$ } & Rod's length & 0.1\tabularnewline
\hline 
\foreignlanguage{english}{$v_{\text{prop}}$ } & Propulsion speed & 0.3\tabularnewline
\hline 
\foreignlanguage{english}{$f_{\text{nem}}$ } & \foreignlanguage{english}{Force of nematodes } & 0.01\tabularnewline
\hline 
\foreignlanguage{english}{$m$ } & Number of nematode nodes & 3\tabularnewline
\hline 
\foreignlanguage{english}{$r$ } & Interaction distance & 0.1\tabularnewline
\hline 
\foreignlanguage{english}{$\xi_{r}$ } & Interaction force scaling & 0.1\tabularnewline
\hline 
\foreignlanguage{english}{$\xi_{\theta}$ } & Interaction torque scaling & 0.1\tabularnewline
\hline 
\end{tabular*}
\end{ruledtabular}

\caption{Model parameters and the non-dimensional values used in the simulations.}

\end{table}

\subsection*{Distribution of active forces along an elastic curve}

The purpose of this appendix is to show that uniform distribution
of active forces minimizes the elastic energy of the elastic deformable
curve. Specifically, describe the deflection of an elastic curve by
$u(x)$ which is circular at equilibrium. Let $x$ be the parameterization
of the curve, so assume that $u(x)$ is periodic with respect to $x$.
Consider the following ``elastic'' energy 
\begin{eqnarray*}
 &  & \mathcal{E}[u,x_{1},...,x_{N}]=\dfrac{K}{2}\int\limits^{1}_{0}|u_{x}|^{2}\text{d}x
\end{eqnarray*}
for periodic $H^{1}(0,1)$ functions $u(x)$ solving the following
differential equation 
\[
-Ku''=\dfrac{f_{p}}{N}\sum\limits^{N}_{i=1}\delta(x-x_{i})-f_{p},
\]
for given points $\{x_{i}\}^{N}_{i=1}\subset[0,1)$. The energy functional
$\mathcal{E}$ accounts for surface tension along the one-dimensional
curve. This functional can be complemented by higher-order terms,
such as a bending rigidity energy. As pointed out in Discussion, at
the length scale under consideration, bending rigidity can be neglected.
For the sake of simplicity, set $K=1$ and $f_{p}=1$. The first term
in the right-hand side describes active forces. The last term is to
guarantee that the solution exists (up to translation). Physically,
this term can be understood as the one due to the volume-preserving
force.

Note that if $u$ is the solution of this equation, then 
\begin{eqnarray*}
 &  & \mathcal{E}[u,x_{1},...,x_{N}]=\dfrac{1}{2N}\sum\limits^{N}_{i=1}u(x_{i})-\dfrac{1}{2}\int\limits^{1}_{0}u(x)\,\text{d}x.
\end{eqnarray*}

The solution $u(x)$ is piecewise quadratic 
\[
u(x)=\dfrac{1}{2}x^{2}+ax+b
\]
with jumps at $x_{i}$ such that 
\[
u'(x^{+}_{i})-u'(x^{-}_{i})=-\dfrac{1}{N}.
\]
The solution can be written with the Green's function given by 
\[
G(x)=\dfrac{1}{2}\left(d(x)^{2}-d(x)+\dfrac{1}{6}\right),
\]
where $d(x)=x\text{ mod }1$. That is, $-G''(x)=\delta(x)-1$. Then
(up to a constant) 
\begin{eqnarray}
\left\{ \begin{array}{rcl}
u & = & \dfrac{1}{N}\sum\limits^{N}_{i=1}G(x-x_{i}),\\
\mathcal{E} & = & \dfrac{1}{2N^{2}}\sum\limits^{N}_{i,j=1}G(x_{i}-x_{j}).
\end{array}\right.\label{energy}
\end{eqnarray}

Without loss of generality, $x_{1}=0$, and denote 
\begin{eqnarray*}
 & \ell_{1}=x_{2}-x_{1},\\
 & \ell_{2}=x_{3}-x_{2},\\
 & ...\\
 & \ell_{N-1}=x_{N}-x_{N-1},\\
 & \ell_{N}=1-x_{N}.
\end{eqnarray*}
Then, using the energy expression in (\ref{energy}), that $G$ is
even, and $\sum\limits^{N}_{i=1}\ell_{i}=1$ we obtain 
\begin{eqnarray*}
\mathcal{E} & = & \dfrac{1}{2N^{2}}\sum\limits^{N}_{i=1}\sum\limits^{N-1}_{k=1}G(s^{(k)}_{i})+\dfrac{1}{2N}G(0),
\end{eqnarray*}
where $s^{(k)}_{i}=d(x_{i+k}-x_{i})\in(0,1)$ where the indices are
understood modulo $N$. Equivalently, $s^{(k)}_{i}=\ell_{i}+...\ell_{i+l-1}$.
Using that $G$ is strictly convex, $\sum\limits^{N}_{i=1}s^{(k)}_{i}=k$,
and Jensen's inequality we derive that 
\[
\dfrac{1}{N}\sum\limits^{N}_{i=1}G(s^{(k)}_{i})\geq G\left(\dfrac{1}{N}\sum\limits^{N}_{i=1}s^{(k)}_{i}\right)=G\left(\dfrac{k}{N}\right)
\]
and thus $\mathcal{E}\geq\mathcal{E}_{\text{eq}}$, where $\mathcal{E}_{\text{eq}}$
is the energy for $x_{i}=(i-1)/N$, and the equality is possible only
for $x_{i}=(i-1)/N$. So the equally spaced configuration minimizes
$\mathcal{E}$ (up to translation).
\end{document}